\documentclass[10pt,twocolumn,letterpaper]{article}

\usepackage{3dv}
\usepackage{times}
\usepackage{epsfig}
\usepackage{graphicx}
\usepackage{amsmath}
\usepackage{amssymb}

\usepackage[pagebackref=true,breaklinks=true,colorlinks,bookmarks=false]{hyperref}

\threedvfinalcopy 

\def\threedvPaperID{299} 

\ifthreedvfinal\fi
\begin{document}

\title{View Correspondence Network for Implicit Light Field Representation}

\author{Süleyman Aslan\\
University of Maryland\\
{\tt\small aslan@umd.edu}
\and
Brandon Yushan Feng\\
University of Maryland\\
{\tt\small yfeng97@umd.edu}
\and
Amitabh Varshney\\
University of Maryland\\
{\tt\small varshney@umd.edu}
}

\maketitle

\begin{abstract}
   We present a novel technique for implicit neural representation of light fields at continuously defined viewpoints with high quality and fidelity. Our implicit neural representation maps 4D coordinates defining two-plane parameterization of the light fields to the corresponding color values. We leverage periodic activations to achieve high expressivity and accurate reconstruction for complex data manifolds while keeping low storage and inference time requirements. However, na\"{i}vely trained non-3D structured networks do not adequately satisfy the multi-view consistency; instead, they perform alpha blending of nearby viewpoints. In contrast, our View Correspondence Network, or VICON, leverages stereo matching, optimization by automatic differentiation with respect to the input space, and multi-view pixel correspondence to provide a novel implicit representation of the light fields faithful to the novel views that are unseen during the training. Experimental results show VICON superior to the state-of-the-art non-3D implicit light field representations both qualitatively and quantitatively. Moreover, our implicit representation captures a larger field of view (FoV), surpassing the extent of the observable scene by the cameras of the ground truth renderings.
\end{abstract}

\section{Introduction}

\begin{figure*}
    \centering
    \includegraphics[width=1.0\linewidth,keepaspectratio]{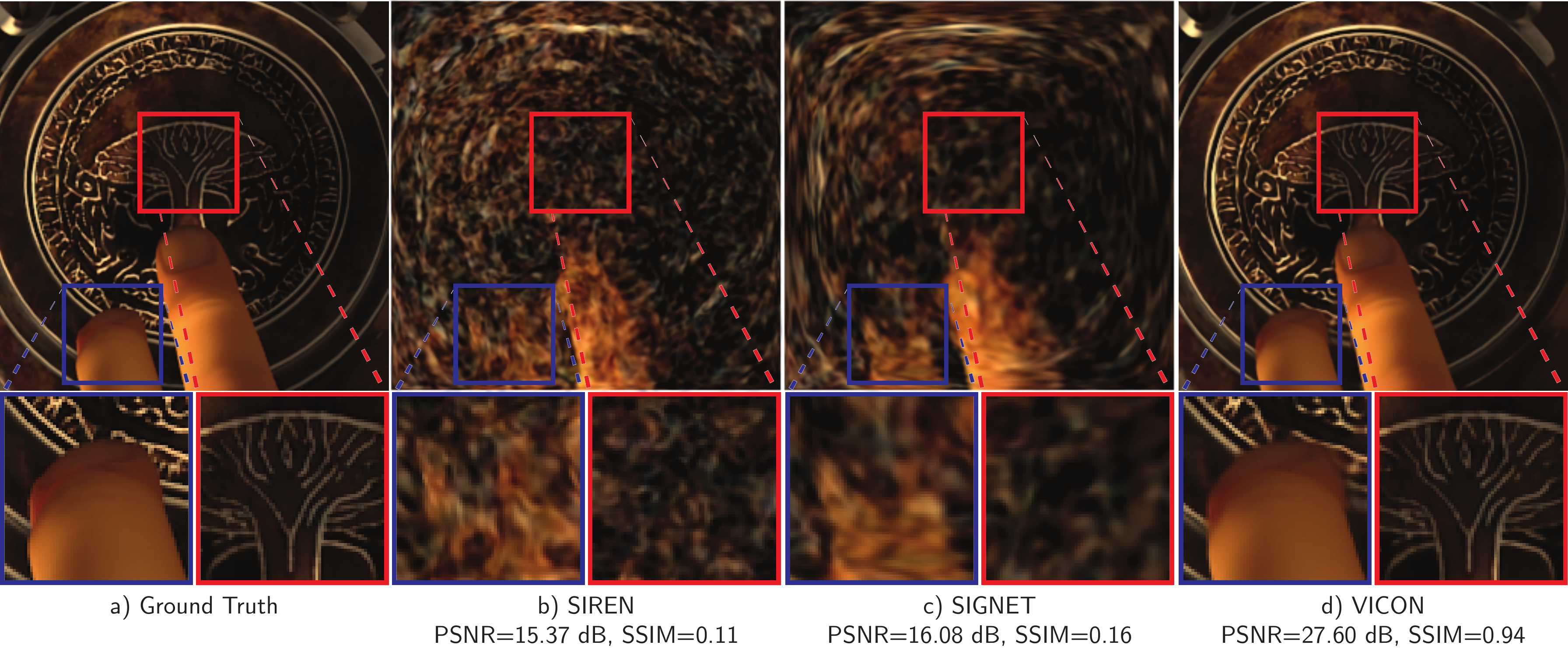}
    \caption{\textbf{Comparison of Novel View Representation.} SIREN (b) and SIGNET (c) fail to accurately represent a novel view due to significant parallax between neighboring viewpoints in the ``shaman\_b\_1'' scene~\cite{kinoshita2021depth}. Our method VICON (d) faithfully renders the novel view and achieves high quality for the fine details of the glyph by effectively incorporating information from source views. All neural networks have the same architecture and the same number of parameters. The hyperparameters of each method are tuned separately to output the best possible representations to each of the methods' capabilities. We include the peak signal-to-noise ratio (PSNR), and structural similarity (SSIM) values as quantitative evaluations.}
    \label{fig:teaser}
\end{figure*}

Light fields are very promising for a variety of interactive 3D graphics applications, especially in conjunction with extended reality (XR) and holographic displays~\cite{overbeck2018system,huang2015light,jones2007rendering,wetzstein2011layered,shi2017near,yamaguchi20163d}. However, excessive storage, transmission, and computation requirements due to the high-dimensionality of the data have hindered light fields from becoming ubiquitous. Several methods have been proposed to compress light fields~\cite{viola2017comparison,wu2017light}.

Implicitly defined continuous representations have emerged as a new alternative for various visual data, including light fields~\cite{sitzmann2020implicit,mildenhall2020nerf,tancik2020fourier,park2019deepsdf}. One of the most prominent representations that does not depend on a ray-based neural renderer is the sinusoidal representation network (SIREN)~\cite{sitzmann2020implicit}. We base our method on SIRENs, and discuss other methods in Section~\ref{sec:relatedwork}. Such a representation is characterized by a series of parameterized, continuous, and differentiable functions. It is parameterized by a neural network trained to map the input coordinates to obtain the corresponding property, such as the color. The defining property of SIREN is the use of periodic activation functions. With appropriate initialization schemes, SIREN can accurately represent complex signals when networks with conventional activation functions fail to preserve finer details. More recently, SIGNET~\cite{feng2021signet}, a derivative of SIREN, provides an efficient and effective implicit light field representation that enables compact storage and super-resolution by using an input transformation strategy based on Gegenbauer polynomials. However, we observe that these sinusoidal representations have a weakness that can severely hinder the performance under challenging conditions, as discussed below.

We consider the light field as a simplified version of the plenoptic function~\cite{adelson1991plenoptic,mcmillan1995plenoptic}, consisting of 4D coordinates as the two-plane parameterization~\cite{levoy1996light,gortler1996lumigraph}, as $L(u,v,x,y)$ where $u,v \in \mathbb{R}$ are the angular coordinates and $x,y \in \mathbb{R}$ are the spatial coordinates of a pixel. In general, to implicitly represent a light field with a continuous and parameterized function, a sinusoidal representation network can be defined as:
\begin{equation}
    \mathcal{F}_\theta(u,v,x,y) = \mathbf{W}_i ( \phi_{i-1} \circ \dots \circ \phi_{1} (u,v,x,y) ) + \mathbf{b}_i
\end{equation}
where $\mathcal{F}$ is the parameterized mapping function for the light field, $\theta$ denotes the parameters of the neural network, and $\phi_i$ is the $i^{th}$-layer of SIREN. As SIRENs are multilayer perceptrons (MLPs), $\phi_i  (\mathbf{x}_i) = \sigma (\mathbf{W}_i \mathbf{x}_i + \mathbf{b}_i)$ with a weight matrix $\mathbf{W}_i$, a bias vector $\mathbf{b}_i$, and a sinusoidal activation $\sigma$.

This network is supervised on the ground-truth values $f(p)$ where $p \in \mathbb{R}^4$ defines the spatio-angular coordinates of a pixel. We minimize an empirical reconstruction loss $\mathcal{L}$, over a dataset $\mathcal{S}$ that contains images at different viewpoints $\mathbf{I}_1,\dots,\mathbf{I}_n$ at known spatio-angular coordinates with RGB colors $f(p_1^{\mathbf{I}_i}),\dots,f(p_{w \times  h}^{\mathbf{I}_i})$, where $w,h \in \mathbb{Z}_{\geq 0}$ denote the spatial resolution. Here, we compute the loss function as the mean absolute error (MAE) between the output of the parameterized mapping function and the ground truth color values:
\begin{equation}
\mathcal{L}_\mathcal{S}(\mathcal{F}_\theta) = \frac{1}{n \times w \times h} \sum_{i}^{n} \sum_{j}^{w \times h} | \mathcal{F}_\theta(p_j^{\mathbf{I}_i}) - f(p_j^{\mathbf{I}_i}) |
\label{eq:ls}
\end{equation}

The optimal parameters $\theta^*$ are obtained by minimizing this loss function. A practical implementation by using an optimizer such as ADAM~\cite{kingma2014adam}, can effectively train the representation network to minimally reduce this loss and provide a high quality reconstruction for any pixel in $\mathcal{S}$.

One of the properties of these representations is the ability to implicitly represent novel viewpoints. Let $\mathcal{S}'$ denote a novel set of views at angular coordinates \emph{unseen} during the training of the sinusoidal representation network. Then, $\mathcal{F}_{\theta^*}(p_j^{\mathbf{I}'_i}),\ \mathbf{I}'_i \in \mathcal{S}',\ j \in [0,w \times h]$ gives us a pixel at a novel view. This is a useful property, which belongs to any continuously defined representation, as it allows super-resolution without any \emph{additional training}. By generalizing this property to the other input dimensions, we can perform super-resolution along any dimension. SIGNET exploits this property to demonstrate the usefulness in various experimental setups and demonstrate up-sampling for the angular dimensions, i.e., representing novel views. The super-resolution is achieved without an explicit supervision, which is different than most super-resolution methods that rely on depth or disparity~\cite{wanner2012globally,wanner2013variational,kalantari2016learning,shi2020learning}, optical flow~\cite{sajjadi2018frame,liu2018learning,wang2020deep}, stereo matching~\cite{jeon2018enhancing,guo2019learning,wang2019learning}, specific deep neural networks~\cite{yoon2015learning,yeung2018fast,gul2018spatial}, or 3D-structured ray-based neural renderers such as NeRF~\cite{mildenhall2020nerf,martin2021nerf,yu2021pixelnerf}.

\begin{figure}[b]
    \centering
    \includegraphics[width=0.85\linewidth,keepaspectratio]{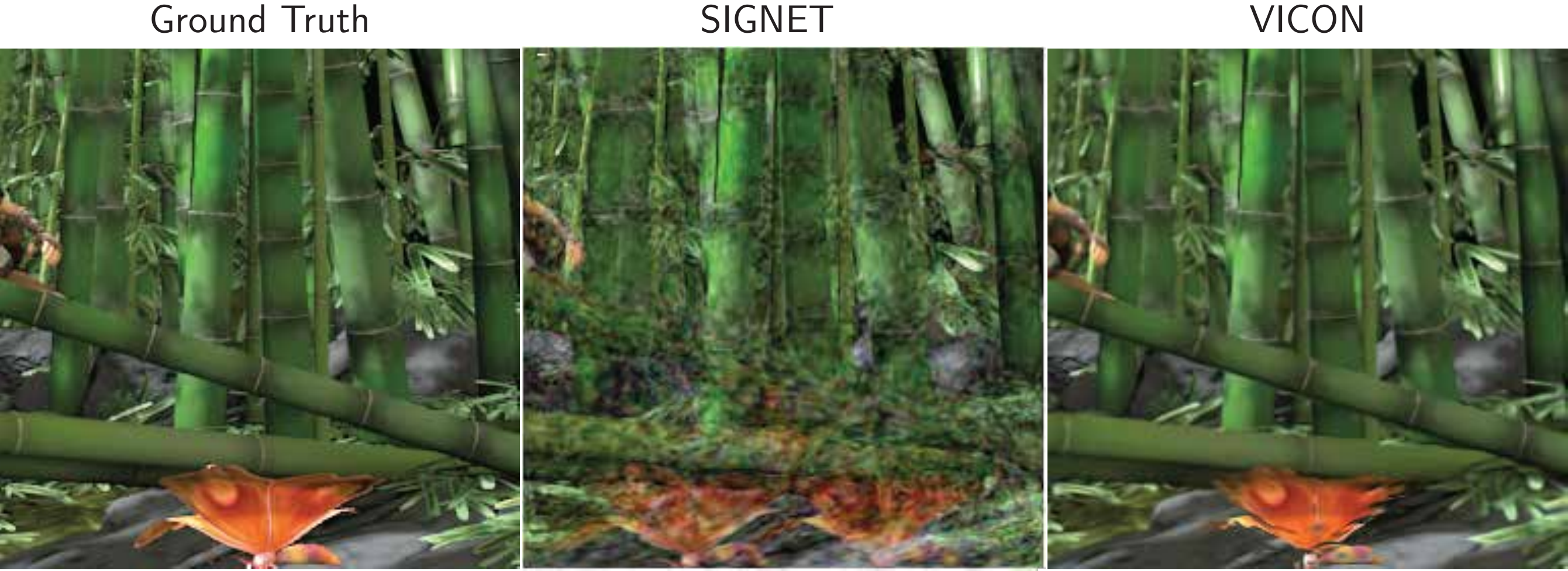}
    \caption{\textbf{ A Common Failure When Interpolating at Novel Views. } Ground truth novel viewpoint (left), rendered novel viewpoint using SIGNET (middle), and rendered novel viewpoint using our method VICON (right). VICON is able to produce visually consistent results when the results produced by SIGNET contain repetitive artifacts. Although SIGNET could accurately encode the training views, the trained network has no knowledge about depth or view correspondence. Therefore, the interpolated result of SIGNET at the novel view point appears to be an incorrect blend of the nearby views.  }
\label{fig:sirennovel}
\end{figure}

\begin{figure*}[t]
    \centering
    \includegraphics[width=1.0\linewidth,keepaspectratio]{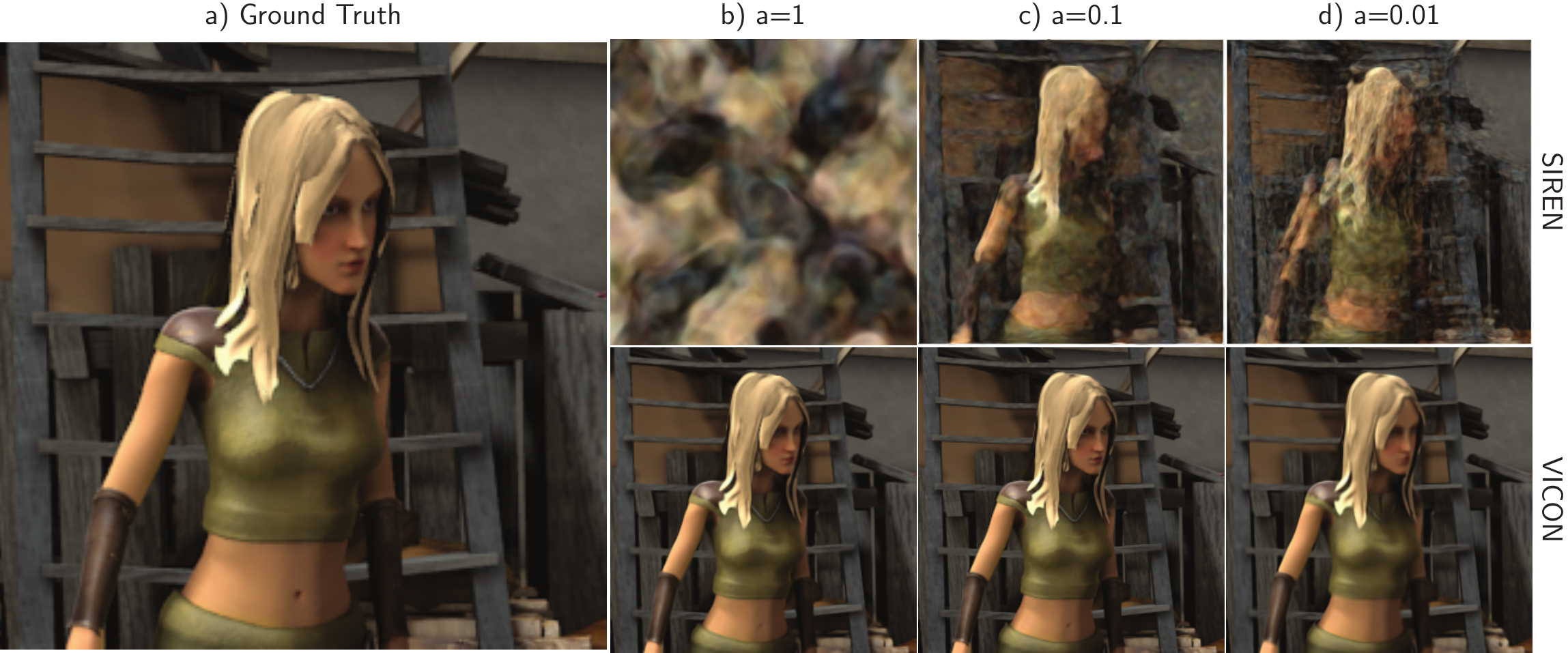}
    \caption{\textbf{The Effect of Range Hyperparameter $a$ on Novel Viewpoints.} Large range values for the angular coordinates prevent representation networks from accurately interpolating between neighboring viewpoints (b, top). Bringing the angular coordinates closer to each other allows networks to interpolate, however the performance is poor (c, top). Using extremely small range values introduces additional noise without improving the results (d, top). In comparison, VICON accurately generates the novel view at any given $a$ without requiring a hyperparameter tuning (bottom).}
\label{fig:range}
\end{figure*}

A current limitation of representations like SIGNET is that they do not constrain or minimize $\mathcal{L}_{\mathcal{S}'}(\mathcal{F}_\theta)$. In other words, the representation network is not conditioned to provide an accurate representation for novel coordinates. Given that $\theta^*$ only depends on $\mathcal{S}$, it is not realistic to expect these parameters to perform well for $\mathcal{S}'$. We perform angular super-resolution with SIGNET using a synthetic light field video dataset~\cite{kinoshita2021depth,sintel}. Results in Fig.~\ref{fig:sirennovel} show that using SIGNET to render novel views perform poorly when the scene content leads to high disparity across different viewpoints. In this paper, we present a training strategy that allows our network to accurately represent novel viewpoints and avoid the shortcomings of SIREN-based representations as shown in Fig.~\ref{fig:sirennovel}. 

Moreover, SIREN introduces an additional hyperparameter $a$, which is set to a default value of 1 and not optimized. This hyperparameter denotes the range of the input space such that the mapping function for the light field is defined as $\mathcal{F}_\theta(u,v,x,y)$, $u,v,x,y \in [-a, a]$. Large $a$ values cause the network to learn the low frequency variation which leads to sparsity, whereas smaller values cause the network to fit the high frequency variation which introduces noisy artifacts.
While spatial coordinates can be populated densely, a lower resolution such as the angular resolution of a light field leads to a sparsity that can affect the performance. As seen in Fig.~\ref{fig:range}, varying this range can mitigate the poor performance. However, even the manually optimized $a$ value does not yield satisfactory results. On the other hand, our method does not require hyperparameter tuning and works well with any $a$.

In this paper, we introduce a novel implicit representation of light fields that approximates the color value of pixels at novel coordinates and enforces the neural network to perform accurate representation at these novel coordinates. We call our network VICON (View Correspondence Network). VICON relies on the pixel correspondence and works well with accurate depth information. More specifically, we make use of disparity and occlusion prediction using stereo pairs to train the representation network when only RGB information is available. Our method does not alter the underlying neural architecture and it can easily adapt to different non-3D structured implicit representation methods without increasing the inference time. Experiments confirm that our method significantly improves the robustness on the representation of novel viewpoints.

In summary, our contributions are as follows:
\begin{enumerate}
    \item We present an implicit neural representation of light fields which faithfully renders high-quality novel views in conditions that challenge the current state-of-the-art methods.
    \item We introduce a coordinate mapping methodology based on pixel correspondence using estimated disparity maps and occlusion masks for multi-view scenes.
    \item We further enhance the estimated disparities during training to improve performance and experimentally show how our approach can also provide a larger field of view (FoV).
\end{enumerate}


\begin{figure*}[t]
    \centering
    \includegraphics[width=1.0\linewidth]{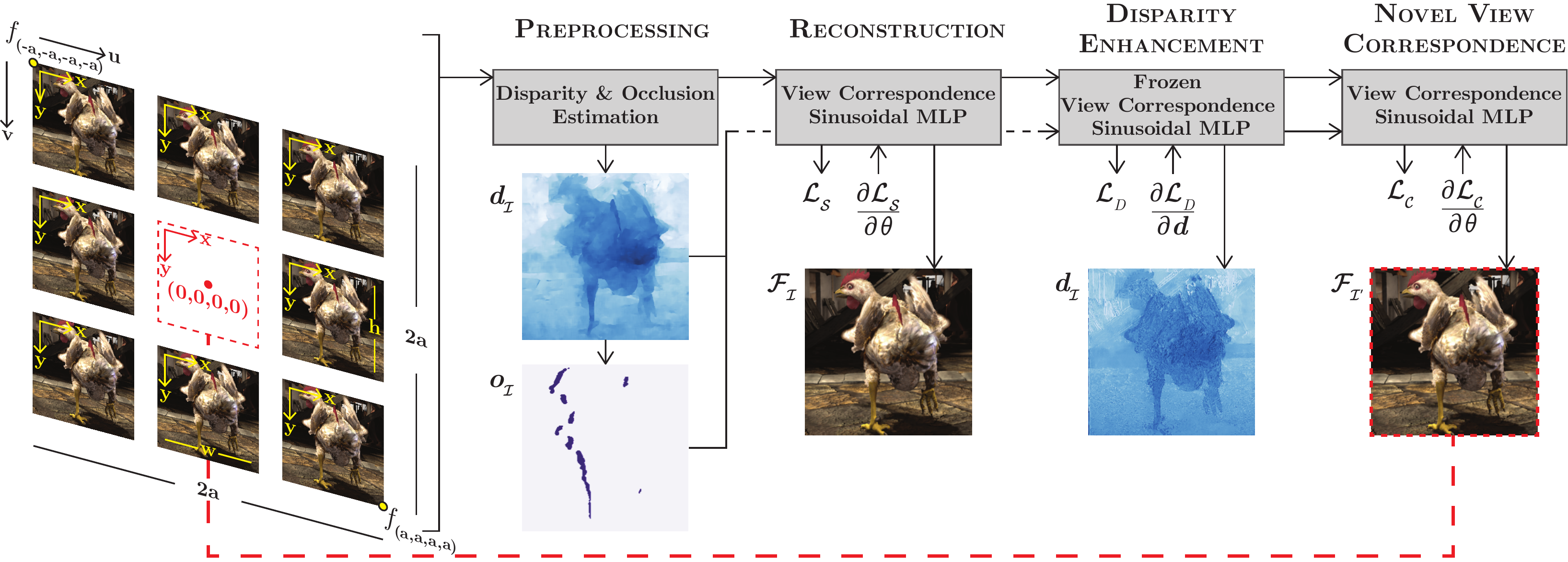}
    \caption{ \textbf{Overall Framework of Our Proposed Method VICON.} Using the multi-view source images $\mathcal{I} \in \mathcal{S}$ and ground truth color values $f$ at the coordinates $(u,v,x,y)$ as input, we estimate the crude disparity maps $d_\mathcal{I}$ and the occlusion masks $o_\mathcal{I}$ with the Stereo Transformer~\cite{li2021revisiting} as a preprocessing step. Then we fit a sinusoidal representation network $\mathcal{F}$ to the known viewpoints based on the loss function $\mathcal{L}_S$ and the gradients with respect to the network parameters $\frac{\partial \mathcal{L}_S}{\partial \theta}$ to perform accurate representation, and also leverage automatic differentiation to enhance the estimated disparity maps by calculating the gradients of the loss function $\mathcal{L}_D$ \emph{with respect to the estimated disparities} $\frac{\partial \mathcal{L}_D}{\partial d}$ to optimize the disparity maps. Finally, we use pixel correspondence based coordinate mapping to compute $\mathcal{L}_C$ and learn to represent the novel viewpoints by optimizing the network using the gradients $\frac{\partial \mathcal{L}_C}{\partial \theta}$. Note that this entire process is end-to-end trainable. Our optimized representation network VICON faithfully represents the novel views of the light field when conventional sinusoidal representation networks fail, and achieves larger field of view (FoV) for novel views compared to the ground truth (see Fig.~\ref{fig:largerfov}). All pixels afford random access without additional computation during inference.}
\label{fig:framework}
\end{figure*}

\section{Methodology} \label{sec:methodology}

Given $n$ known source images $\mathcal{I} \in \mathcal{S}$, we minimize the loss over $m$ additional unknown novel views $\mathcal{I}' \in \mathcal{S}'$ to train VICON. Thus, despite using only a sparse representation of the scene, our network fits to a denser representation. An overall framework depicting the proposed methodology is given in Fig.~\ref{fig:framework}. Note that, although $n$ is limited by the available known viewpoints of the scene, $m$ can be as large as preferred. However, larger values of $m$ impose additional training time and might require neural networks with a larger number of parameters to obtain a good fit. We define the relationship in the angular coordinate space between a novel view and all source images by the following affine combinations:
\begin{equation}
\begin{split}
u_{\mathcal{I}'} = \sum_{i=1}^n \alpha_i u_{\mathcal{I}_i}, \quad v_{\mathcal{I}'} = \sum_{i=1}^n \beta_i v_{\mathcal{I}_i}, \\
\text{s.t.} \quad \sum_i \alpha_i = \sum_i \beta_i = 1
\end{split}
\label{eq:angular_relation}
\end{equation}

Although we can define a non-convex combination, i.e., an \emph{extrapolated view}, we focus on only \emph{interpolated views} and assume that $\alpha_1, \dots, \alpha_N$ and $\beta_1, \dots, \beta_N$ are non-negative scalars. Then, for a given novel view, $\mathcal{I}'$, we first consider the spatial coordinates of pixels $(x,y)$ that are present in the known views and perform a mapping of the novel pixels at these coordinates to their corresponding pairs in source images by shifting them in the spatial coordinate space. To compute the difference in the spatial dimension, we calculate the divergence in the angular coordinate space of the sinusoidal network. Then, we perform appropriate scaling by multiplying by the per-pixel disparity, which we denote as $d(u,v,x,y)$\footnote{We assume that the per-pixel disparity is same for both spatial dimensions, however it can easily be adapted when disparity depends on the dimension by calculating these values separately as $d_x$ and $d_y$.}, and dividing by the spatial resolution and the half of the angular coordinate range\footnote{Pixel positions are non-negative for the images but the representation network models coordinates in both negative and positive regions.} of the representation network. As we know the RGB values of pixels from the known coordinates, we can approximate the color for the novel coordinates using~(\ref{eq:correspondence}),
\begin{equation}
\mathcal{F}_\theta   
    \begin{pmatrix}    
        {u_j }, \\ 
        {v_j }, \\ 
        {x } {+} \frac{{2}{(u_i  - u_j)}{d(u_i,v_i,x,y)}}{{w \times a}}, \\ 
        {y } {+} \frac{{2}{(v_i  - v_j)}{d(u_i,v_i,x,y)}}{{h \times a}}
    \end{pmatrix} 
	\approx
f   \begin{pmatrix}
        u_i,\\ 
        v_i,\\ 
        x,\\ 
        y
    \end{pmatrix}
\label{eq:correspondence}
\end{equation}
where $(u_j,v_j)$ are the angular coordinates of a novel view, and $(u_i,v_i)$ are the angular coordinates of a known view.

\subsection{Handling Edge Cases}

\begin{figure*}
    \centering
    \includegraphics[width=1.0\linewidth,keepaspectratio]{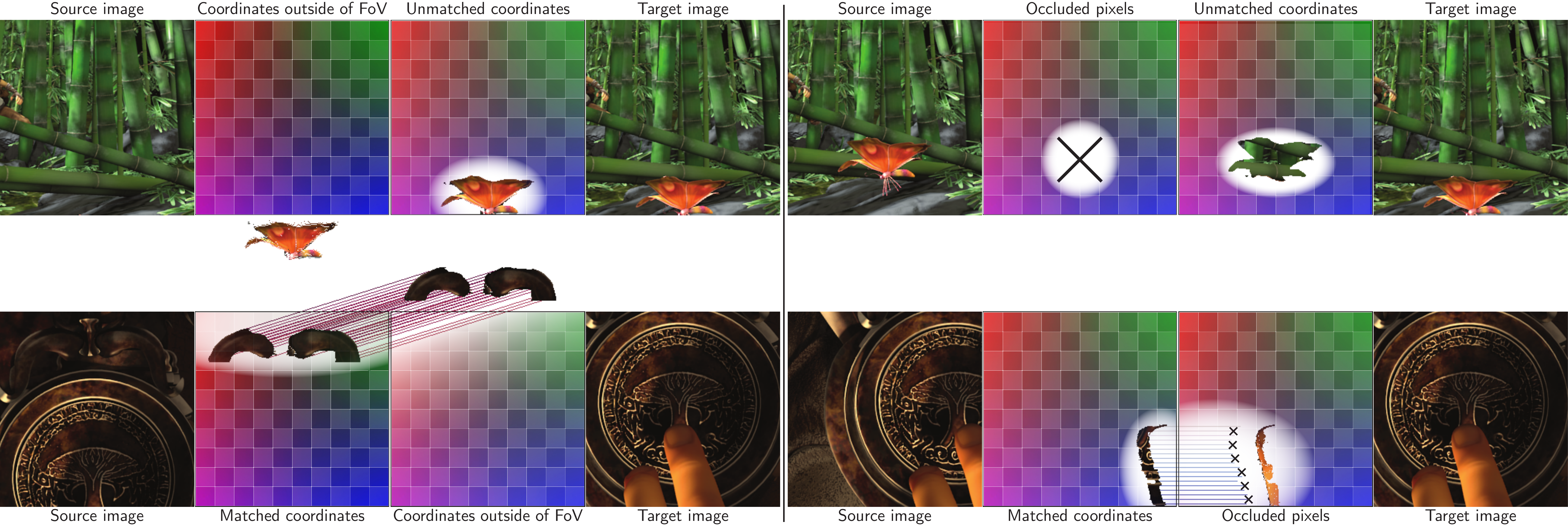}
    \caption{\textbf{Visualization of Different Cases for Correspondence.} Coordinates in the target (novel) image lie outside of the FoV in the source image (top). Pixels in the target view are occluded in the source view (upper middle). Coordinates in the source image move outside of the image plane in the target view (lower middle). Pixels in the source view are occluded in the target view (bottom).}
\label{fig:correspondence}
\end{figure*}

Note that this correspondence does not model the occlusion. Therefore, na\"{i}vely using Equation (\ref{eq:correspondence}) will not be accurate when pixels are visible from some of the source viewpoints but occluded from the novel viewpoint due to the 3D structure of the scene. To further analyze this correspondence, we consider the following cases,
\begin{enumerate}
    \item A coordinate in the novel view lies outside the image frame in a source view. When $|x| > a$ or $|y| > a$, $f(u_i, v_i, x, y)$ is not available, therefore Equation (\ref{eq:correspondence}) will not impose a constraint on the RGB colors for the novel coordinates. If these coordinates can match pixels in other source views, the color values will be optimized using those pixels. Otherwise, the training of the representation network will exclude these coordinates, which default to the SIGNET behavior. 
    \item A coordinate in a source view moves outside of the image frame in the novel view. We map a known color $f(u_i, v_i, x, y)$ to $\mathcal{F}_\theta(u_j, v_j, x', y')$ using (\ref{eq:correspondence}) where $|x'| > a$ or $|y'| > a$. As VICON or any other sinusoidal representation network can still operate on these coordinates, the optimization will be carried out using back-propagation. In fact, this allows us to render the novel view using larger spatial dimensions compared to the dimensions of the target images. We show an example showing this behaviour in Fig.~\ref{fig:largerfov}. 
    \item A pixel at a coordinate in the novel view is occluded in a source view. Similar to the first case, Equation (\ref{eq:correspondence}) will not affect these coordinates because mapping is not available at these coordinates. We aim to learn the color values of the pixels from other views.
    \item A pixel at a coordinate in a source view becomes occluded in the novel view. This is problematic since Equation (\ref{eq:correspondence}) will erroneously match the coordinates, which requires special handling. To solve this problem, we leverage occlusion information and exclude the gradients of pixels occluded in the novel viewpoints from the backpropagation.
\end{enumerate}

We illustrate all of these cases in Fig.~\ref{fig:correspondence}. Accordingly, we define a correspondence loss as:
\begin{equation}
\mathcal{L}_C = \frac{1}{N} \sum_{i}^{N} 1_A(p_i) | \mathcal{F}_\theta(p_i') - f(p_i) |
\label{eq:lc}
\end{equation}
where $p_i$ denotes the spatioangular coordinates of known color values, $p_i'$ is the novel view correspondence given in (\ref{eq:correspondence}) and, $1_A(p_i)$ is the indicator function denoting the occlusion. $1_A(p_i) = 0$, when $p_i'$ is occluded and $1_A(p_i) = 1$, otherwise.

\begin{figure*}[t]
    \centering
    \includegraphics[width=1.0\linewidth,keepaspectratio]{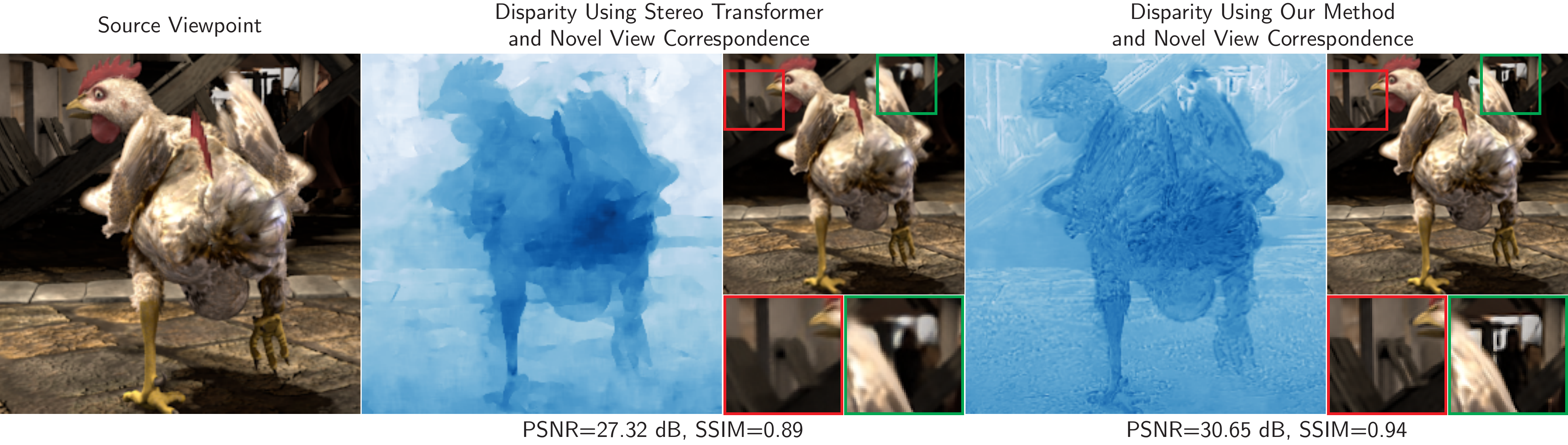}
    \caption{\textbf{Enhancing Estimated Disparity Maps.} RGB image from a source viewpoint (left), disparity estimation  using Stereo Transformer~\cite{li2021revisiting} with the novel viewpoint representation performance (middle), and the improved disparity estimation by using the auto-differentiation capabilities of the functional representation with the novel viewpoint representation performance (right). VICON produces sharp edges, more consistent regions in the disparity map, and accurately learns fine details. Our disparity maps improve the results both qualitatively and quantitatively.}
\label{fig:dispenhance}
\end{figure*}

Assuming only RGB information is available, we aim to obtain accurate disparity maps and occlusion masks to implement Equation (\ref{eq:lc}) by exploiting the multi-view information in the light fields. As the neural representation is differentiable, unlike conventional discrete representations, we design an optimization in the input space to obtain disparity maps by leveraging automatic differentiation of the representation network's outputs with respect to the disparity maps. Although Equation (\ref{eq:correspondence}) is designed to estimate the color values at novel views, we can also use the correspondence between different known viewpoints where one viewpoint is considered as the source and the other as the target. However, this requires $\mathcal{F}$ to be accurate for $\mathcal{I} \in \mathcal{S}$. Therefore, we simultaneously train VICON using the images from known viewpoints without the correspondence given in (\ref{eq:correspondence}), and parameterize $\mathcal{F}$ with the input disparities without updating the network weights $\theta$. To optimize disparities, we calculate the loss slightly modified from Equation (\ref{eq:lc}):
\begin{equation}
\mathcal{L}_D = \frac{1}{N} \sum_{i}^{N} 1_A(p_i) | \mathcal{F}_d(p_i') - f(p_i) |
\label{eq:ld}
\end{equation}
where $p_i'$ is the \emph{known} view correspondence. Another difference between $\mathcal{L}_D$ and $\mathcal{L}_C$ is that we use $\mathcal{L}_D$ to optimize disparity maps with gradients $\frac{\partial \mathcal{L}_D}{\partial d}$, whereas $\mathcal{L}_C$ is used to optimize the network weights with gradients $\frac{\partial \mathcal{L}_C}{\partial \theta}$. This disparity optimization relies significantly on the initialization due to the abundant local minima, therefore we use Stereo Transformer~\cite{li2021revisiting} as a preprocessing step to obtain a proper initialization. We find predicted occlusion masks to be sufficient, therefore we do not further optimize the occlusion information.

Note that although we present and illustrate our approach in a step-by-step manner (Fig.~\ref{fig:framework}), the entire process is end-to-end trainable which is how we implement in our experiments. We use two optimizers: ADAM to optimize network weights and SGD~\cite{robbins1951stochastic} to optimize the disparity maps. We use $\mathcal{L}_D$ for SGD and combine loss functions in (\ref{eq:ls}) and (\ref{eq:lc}) for ADAM:
\begin{equation}
\mathcal{L}_{total} = \lambda_S \mathcal{L}_S + \lambda_C \mathcal{L}_C
\end{equation}
where $\lambda_S, \lambda_C$ are the weights of the respective loss functions. 


\section{Experimental Results}

We use the 4D light field video dataset provided in~\cite{kinoshita2021depth} to evaluate VICON. This dataset has considerably larger parallax compared to other datasets such as the Stanford Light Field dataset~\cite{vaish2008new} or Real Forward-Facing dataset~\cite{mildenhall2019local}, which provides a challenging setup for encoding novel viewpoints. Since we build upon the idea of implicitly performing angular up-sampling as shown with SIGNET and do not change the underlying neural architecture or impose additional computation for the inference, we compare VICON to SIREN and SIGNET. Note that we do not compare against 3D-structured representations such as SRN~\cite{sitzmann2019scene} and NeRF~\cite{mildenhall2020nerf} as they leverage information based on the ray direction and render images with a ray-based neural renderer which is orthogonal to our way of approaching the implicit neural representations. In contrast, SIREN, SIGNET, and our method do not have any assumptions with the input data, resulting in lightweight and efficient representations. The dataset in our experiments contains $24$ synthetic 4D light fields with $9$ viewpoints in both angular dimensions and are rendered with $1204 \times 436$ pixels.  We use $7$ scenes with $256 \times 256$ cropped images to fit the neural networks. Although the dataset provides ground truth disparity values obtained by transforming the depth values, we only use the RGB information and obtain disparity maps as discussed in Section~\ref{sec:methodology}. 
For VICON, we set the weights as $\lambda_S = \frac{n}{n+m}$ and $\lambda_C = \frac{m}{n+m}$ to promote a balanced fitting of source and novel views. For SIGNET, we select Gegenbauer's $\alpha$ as $0.5$ as suggested by the authors and select network input size C as $128$. 
All networks are sinusoidal MLPs with 5 linear layers and sine activations. We initialize the weights according to the initialization scheme discussed in~\cite{sitzmann2020implicit}. 
All hidden layers have $1024$ neurons. We use a learning rate of $10^{-4}$ for ADAM and $10^{3}$ for SGD without momentum. Additionally, we decay the learning rate by $\gamma = 0.9$ every $150$ steps.

\begin{table}[t]
    \centering
    \setlength{\tabcolsep}{1.75pt}
    \label{tab:quantitative}
    \begin{tabular}{lcc cc cc}
    \hline
    & \multicolumn{2}{c}{SIREN} & \multicolumn{2}{c}{SIGNET} & \multicolumn{2}{c}{VICON} \\
    4D Scenes  & PSNR & SSIM & PSNR & SSIM & PSNR$\uparrow$ & SSIM$\uparrow$ \\ \hline
    ambushfight\_5  & 18.54 & 0.32 & 24.68 & 0.59 & \textbf{35.88} & \textbf{0.94}  \\ 
    bamboo\_2       & 18.64 & 0.44 & 19.31 & 0.46 & \textbf{33.55} & \textbf{0.96}  \\ 
    chickenrun\_3   & 22.22 & 0.55 & 24.49 & 0.70 & \textbf{30.92} & \textbf{0.94}  \\ 
    foggyrocks\_2   & 18.32 & 0.29 & 29.31 & 0.72 & \textbf{33.03} & \textbf{0.96}  \\ 
    questbegins\_1  & 27.13 & 0.70 & 30.78 & 0.75 & \textbf{42.40} & \textbf{0.97}  \\ 
    shaman\_b\_1    & 15.37 & 0.11 & 16.08 & 0.16 & \textbf{27.60} & \textbf{0.94}  \\ 
    thebigfight\_2  & 15.72 & 0.11 & 24.22 & 0.76 & \textbf{30.55} & \textbf{0.94}  \\  \hline
    Mean            & 19.42 & 0.36 & 24.12 & 0.59 & \textbf{33.41} & \textbf{0.95} \\
    \hline
    \end{tabular}
    \vspace{5pt}
    \caption{\textbf{Quantitative Evaluations.} VICON consistently produces high quality results across multiple scenes, while the performance of SIREN and SIGNET is unstable and overall achieve a lower quality than VICON.}
\end{table}

\begin{figure*}
\begin{center}
    \includegraphics[width=1.0\linewidth,keepaspectratio]{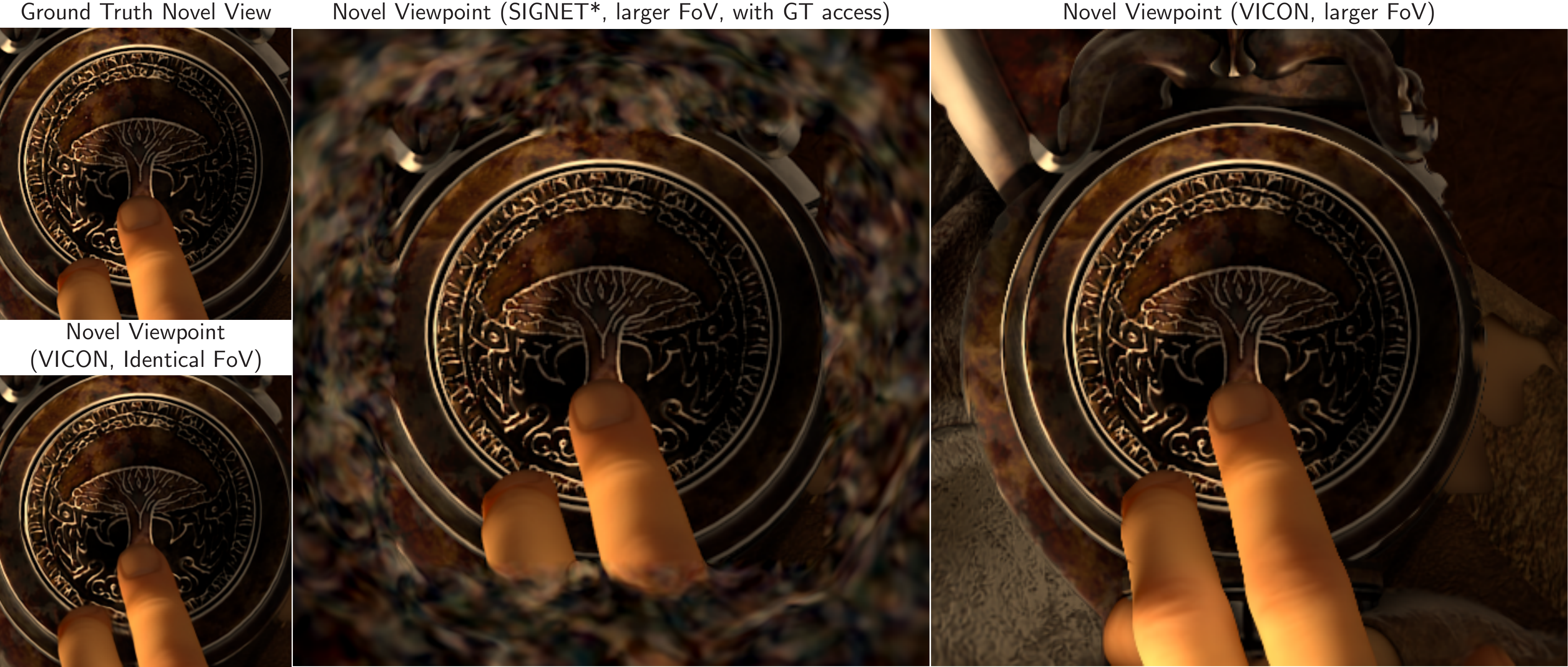}
\end{center}
    \caption{\textbf{Novel Viewpoints with FoV Extrapolation.} Ground truth unseen novel view (top left), rendering of the novel view using VICON with identical FoV (bottom left), rendering of the novel view using SIGNET~\emph{having access to the ground truth data} with a larger extrapolated FoV (middle), and rendering of the novel view using VICON~\emph{without access to the ground truth data} and with a larger extrapolated FoV (right). Our method is able to effectively extrapolate the pixels based on the available multiview information thanks to the disparity-aware training strategy described in Section~\ref{sec:methodology}. Note that SIGNET is not able to represent the novel view accurately when using only the source viewpoints (see Fig.~\ref{fig:teaser}).}
\label{fig:largerfov}
\end{figure*}

We present quantitative results on the performance of representing novel views in Table~\ref{tab:quantitative}. Depending on the mean disparity of the scenes, SIREN and SIGNET show varying levels of performance ranging between $15-30$ dB PSNR and $0.1-0.7$ SSIM. In addition to inconsistent performance, the novel view renderings are qualitatively unsatisfactory (see Figures ~\ref{fig:teaser}, ~\ref{fig:sirennovel}, and ~\ref{fig:range} for examples). In comparison, VICON consistently produces high quality outputs irrespective of the magnitude of the disparity and outperforms the state-of-the-art implicit neural representation methods.

We also evaluate our disparity optimization method used to train VICON as discussed in Section~\ref{sec:methodology}. We separately fit two VICONs using unoptimized and optimized disparity maps and compare the results. We observe that, for the ``chickenrun\_1'' scene, VICON with unoptimized disparity maps achieves a PSNR of $27.32$ dB and an SSIM of $0.89$ whereas VICON with the disparity maps using our method improves the performance to a PSNR of $30.65$ dB and an SSIM of $0.94$. Moreover, as seen in Fig.~\ref{fig:dispenhance}, the qualitative comparison of the estimated disparity maps demonstrates that our method leads to sharper boundaries, reveals finer details, and obtains more consistent regions. Also, use of unoptimized disparities results in blurrier rendering for the novel views.

Finally, we show the capability of VICON to achieve a larger FoV for the novel view compared to the ground truth. Case (b) of the correspondence explained in Section~\ref{sec:methodology} enables VICON to fit to the pixels outside the image frame. We leverage this property by inputting a larger area in the coordinate space to the network  to render the novel view using larger spatial dimensions. Fig.~\ref{fig:largerfov} shows that VICON seamlessly and accurately extrapolates the pixels by using the multi-view information. We also use a SIGNET to perform extrapolation as comparison. Since SIGNET is not able to represent the novel view accurately, we let SIGNET have access to the ground truth data for this evaluation in order to provide a comparison. It is challenging for SIGNET to perform image extrapolation as the multi-view information is unused.

\section{Related Work} \label{sec:relatedwork}
Implicit coordinate-based neural representations have become increasingly popular in 3D graphics and vision in recent years.
These representations can be generally divided into two groups: 1) 3D-structured representations with volume rendering and ray-based neural renderers, and 2) light field representations without volume rendering.
In this section, we provide a brief overview on these two lines of work and discuss their connections to our method.

\subsection{3D-structured Representations}
NeRF~\cite{mildenhall2020nerf} and its variants have made pioneering progress in producing photorealistic renderings of a 3D scene from just a handful of real-world images.
While the task of novel view synthesis through 3D reconstruction has been studied for decades, the introduction of Neural Radiance Fields (NeRF) has been pivotal in initiating the current trend of \textit{using neural networks} to produce 3D scene properties such as radiance and opacity at a location parametrized by its 3D coordinates.

Most recently, the success of Plenoxels~\cite{yu2021plenoxels} shows that the setup of differentiable volumetric rendering alone is sufficient to achieve state-of-the-art novel view synthesis results, without relying on any neural network to predict radiance and opacity as in the original NeRF.

Although the volumetric rendering setup enables high-quality rendering, it does not come without costs.
While some methods seek to accelerate the rendering of the scene representation learned with NeRF, they alleviate NeRF's rendering time complexity only by sacrificing its space complexity.
To achieve real-time rendering, these methods generally resort to pre-computing certain parts of the scene representation and storing them in cache, essentially reducing the inefficiency of neural network inference to the traditional scenario of volume rendering.
However, these methods could easily bloat the space complexity of the scene representation from a few megabytes to several gigabytes, making them unwieldy for mobile applications.
Our method differs from these methods as we do not model 3D volumetric radiance.
Our method falls into the alternative category of light field representations which will be detailed below.

\subsection{Light Field Representations}
Representations such as SIGNET and LFN~\cite{sitzmann2021light} have been proposed as an alternative to NeRF for coordinate-based neural representations.
Unlike NeRF, which uses alpha-composite-based volumetric rendering and thus focuses on radiance and opacity at each 3D location, these representations directly map an oriented camera ray to the accumulated radiance observed by that ray.
This strategy bypasses the need to evaluate multiple samples to render a single pixel - one pixel corresponds to one neural network pass.
As a result, these representations accelerate rendering and reduce memory consumption by orders of magnitude compared to the previous methods.

However, a crucial limitation of light field representations is the explicit multi-view inconsistency.
While NeRF-based methods are inherently multi-view consistent because they directly model the 3D scene structure, light field representations model the scene from the surface appearance and lack an explicit understanding of the scene geometry.
In our approach, we induce geometric information to VICON by supervising its ray-based representation with disparity information.
Since our techniques are applied during training, we avoid incurring a higher time or space complexity during rendering.

\section{Limitation and Discussion}
With VICON, we can model the cameras in the $(u,v)$ plane as long as there are pixels in the image plane of the camera to be inferred from neighbors. Similarly, the scene points that can be rendered by VICON are limited to the points where the ray from the camera to the point passes through the image plane of the camera. 

Another topic not covered in this paper is light field depth estimation~\cite{wang2015occlusion,shin2018epinet}. Our representation is dependent on the accuracy of the disparity estimation, and since our method performs pairwise stereo estimate, it ignores estimating depth or disparity from the light field as a whole.
However, it would be interesting to investigate if our approach would similarly enhance the depth produced by light field depth estimation methods.

Moreover, like other methods that compute view correspondence through color consistency between different viewpoints, our method would not naturally handle view-dependent effects like reflection and shadows.
A worthwhile future direction would be to examine if the spherical-harmonics-based techniques introduced by previous works~\cite{wizadwongsa2021nex,yu2021plenoxels,mildenhall2020nerf} to model view-dependent effects would improve the robustness of the disparity maps produced by VICON.

\section{Conclusion}
We present VICON, a novel implicit representation of light fields that accurately renders novel views under challenging conditions. 
We leverage stereo matching, multi-view correspondence and automatic differentiation to obtain precise disparity and occlusion information to enforce constraints for the representation of novel coordinates. 
Experimental results demonstrate the superiority of VICON over the state-of-the-art in more robustly performing novel view representation when the training views contain large disparities.
The success of VICON indicates that the emerging trend of implicit neural representations could significantly benefit from well-established computer vision techniques like stereo disparity estimation.
We hope that VICON can inspire further research into improving the practicality and usefulness of implicit neural representations in light field processing pipelines.

{\small
\bibliographystyle{ieee_fullname}
\bibliography{egbib}
}

\end{document}